\documentclass[prl,amssymb,amsmath,amsfonts,superscriptaddress,twocolumn,showpacs,reprint]{revtex4-1}

\usepackage{graphicx}
\usepackage{color} 
\usepackage{hyperref} 
\renewcommand{\vec}[1]{\mathbf{#1}}
\renewcommand{\d}{\mathrm{d}}

\usepackage{textcase}
\usepackage{titlesec}
\makeatletter
\newcommand*{\balancecolsandclearpage}{%
  \close@column@grid
  \clearpage
  \twocolumngrid
}
\makeatother
\usepackage[final]{pdfpages}

\begin{document}

\title{Emergent spatial structures in flocking models: a dynamical system insight}
\author{Jean-Baptiste Caussin}
\affiliation{Laboratoire de Physique de l'Ecole Normale Sup\'erieure de Lyon, Universit\'e de Lyon, CNRS, 46, all\'ee d'Italie, 69007 Lyon, France}
\author{Alexandre Solon} 
\affiliation{Universit\'e Paris Diderot, Sorbonne Paris Cit\'e, MSC, CNRS, 75205 Paris, France}
\author{Anton Peshkov} 
\affiliation{LPTMC, CNRS, Universit\'e Pierre et Marie Curie, 75252 Paris, France}

\author{Hugues Chat\'e} 
\affiliation{Service de Physique de l'Etat Condens\'e, CEA-Saclay, CNRS, 91191 Gif-sur-Yvettes, France}
\affiliation{Max Planck Institute for Physics of Complex Systems, N\"othnitzer Stra\ss e 38, 01187 Dresden, Germany}
\affiliation{LPTMC, CNRS, Universit\'e Pierre et Marie Curie, 75252 Paris, France}

\author{Thierry Dauxois}
\affiliation{Laboratoire de Physique de l'Ecole Normale Sup\'erieure de Lyon, Universit\'e de Lyon, CNRS, 46, all\'ee d'Italie, 69007 Lyon, France}

\author{Julien Tailleur} 
\affiliation{Universit\'e Paris Diderot, Sorbonne Paris Cit\'e, MSC, CNRS, 75205 Paris, France}

\author{Vincenzo Vitelli}
\affiliation{Instituut-Lorenz for Theoretical Physics, Universiteit Leiden, 2300 RA Leiden, The Netherlands}
\author{Denis Bartolo}
\affiliation{Laboratoire de Physique de l'Ecole Normale Sup\'erieure de Lyon, Universit\'e de Lyon, CNRS, 46, all\'ee d'Italie, 69007 Lyon, France}

\begin{abstract}
We show that hydrodynamic theories of polar active matter 
generically possess inhomogeneous traveling solutions. We introduce a unifying dynamical-system framework to establish the shape of  these intrinsically nonlinear patterns, and  show that they  correspond to those  hitherto observed in experiments and numerical simulation:  periodic density waves, and solitonic bands, or polar-liquid droplets both cruising in isotropic phases. We elucidate their respective multiplicity and mutual relations, as well as their existence domain.
\end{abstract}

\maketitle

\begin{figure}
\begin{center}
\includegraphics[width=\columnwidth]{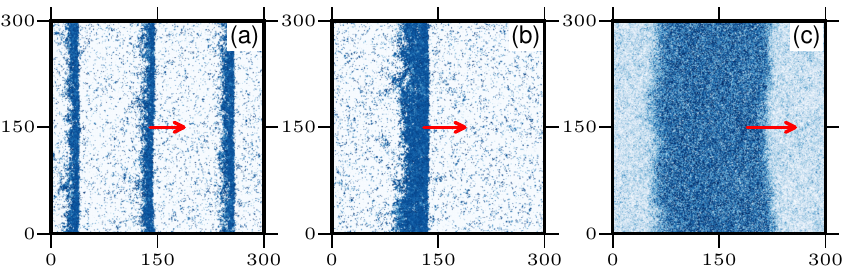}
\caption{Band patterns observed in agent-based
  simulations. (a)~Smectic arrangement of polar bands in the Vicsek
  model with vectorial noise~\cite{ChatePRE}. Speed $v=0.5$, noise
  intensity $\eta=0.6$ and density $\rho=1.1$. (b)~A solitary band
  observed for the same model and same parameters as in (a).
  (c)~Polar-liquid droplet in an isotropic phase observed in the active Ising
  model~\cite{spins}, with inverse temperature $\beta=6$, density
  $\rho=5$, hopping rate $D=1$, and bias $\epsilon=0.9$. More simulation details can be found in~\cite{supp}.}
\label{fig1}
\end{center}
\end{figure}

Could the emergence of collective motion in fish schools, bird flocks,
and insect swarms be understood within a unified physical framework?
A growing stream of works has approached this provocative question following the seminal work
of Vicsek {\it et al.}, who  considered
self-propelled point particles interacting solely via local velocity-alignment rules~\cite{Vicsek}. This  model displays a spontaneous rotational-symmetry breaking leading to orientational order~\cite{Vicsek,Vicsek_review,Marchetti_review}.  
In addition, a number of subsequent 
simulations and experiments
have revealed an even more surprising feature.
At the onset of collective motion, 
 despite the lack of any attractive interactions, polar active matter
self-organizes in the form of   band-shape
swarms ~\cite{ChatePRL,ChatePRE,MarchettiPRE,BaskaranPRE,Bertin,spins,Bausch,rollers,Ihle2013}.  
However, depending on the specifics of the systems, these dynamical patterns take three different  forms: (i)   delocalized density waves~\cite{Bausch,ChatePRE}, as exemplified in Fig.~\ref{fig1}a, (ii) solitonic structures~\cite{Bertin,Ihle2013,rollers},  Fig.~\ref{fig1}b, and (iii) phase-separated
states~\cite{MarchettiPRE,BaskaranPRE,spins},  Fig.~\ref{fig1}c.
Although it is now clear that they are responsible for the first-order nature of the transition towards collective motion~\cite{ChatePRL,spins,Ihle2013}, no  unifying theory exists to account for the origin and the variety of these band patterns. 

In this letter, we convey a comprehensive description of  the propagative excitations of polar active matter. Using a hydrodynamic description and dynamical system concepts, we establish the shape of  these intrinsically nonlinear band structures, and  show that they  correspond to those  observed in all the available experiments and numerical simulation, see e.g.~\cite{ChatePRE,MarchettiPRE,BaskaranPRE,Bertin,spins,Bausch,rollers,Ihle2013,Weber2013,Pawel2012}.

\begin{figure}
\includegraphics[width=\columnwidth]{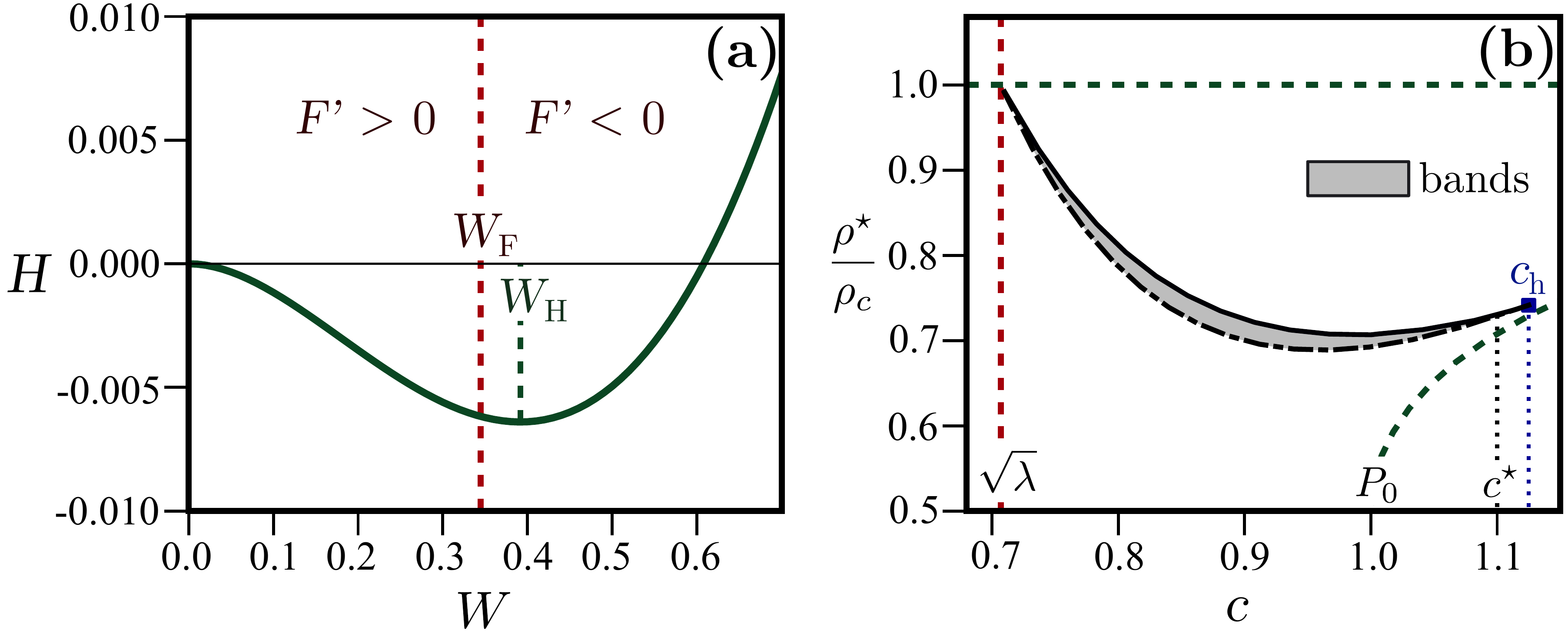}
\caption{(a)~Sketch of the motion of an oscillating point particle in the  effective potential $H(W)$ for $P_0 = 1$, $\rho_{\rm c} = 1$,
  $\lambda = 0.5$, $\xi = 1$, $c = 0.9$ and $\rho^\star = 0.7$. The
  system looses or gains energy $H$ when $F'(W)>0$ or
  $F'(W)<0$. (b)~Band phase diagram for the same parameters. Non-linear bands  exist only in the grey
  region.  The dashed lines correspond the the conditions  $W_H>0$, and $W_F>0$. For $c<c^*$, the black full line corresponds to the
  supercritical Hopf-bifurcation. Polar-liquid droplets  states are observed only at  $c=c_h$. }
\label{fig2}
\end{figure}

Our starting point is a  
hydrodynamic description of 
compressible polar active fluids~\cite{Toner,Marchetti_review}. Since we are 
chiefly interested  in structures varying only along the main direction of motion, 
we focus here on a  one-dimensional problem.
The local density field $\rho(x,t)$ obeys a conservation equation which complements 
the equation governing the momentum field $W(x,t)=\rho(x,t)P(x,t)$, where $P(x,t)\in [0,1]$ is
a polarization field. Following Toner and Tu~\cite{Toner} these equations read
\begin{align}
\label{hydro1}	\partial_t \rho + \partial_x W &= 0\\
\label{hydro2}	\partial_t W +\xi W\partial_x W &= a_2 W -a_4 W^3 
 -\lambda \partial_x\rho +D\,\partial_{xx}W 
\end{align}
where all the coefficients  {\it a priori} depend  both on
$\rho$ and $W^2$. 
These phenomenological equations were introduced  to account for a continuous mean-field transition from a homogeneous isotropic state with $\rho=\rho_0$ and $P=0$ when $a_2<0$, to a homogeneous polarized state with $P=\rho_0^{-1}\sqrt{a_2/a_4}$ when $a_2>0$. In addition, the $\lambda$ term reflects the pressure gradient induced by density heterogenities. $\xi$ and $D$ are two transport coefficients associated respectively with the advection and  the diffusion of the local order parameter.
 Following~\cite{Bertin}, we now look for
propagating solutions of Eqs.~\eqref{hydro1} and \eqref{hydro2}: $\rho
= \rho(x-ct)$ and $W = W(x-ct)$, where $c$ is the propagation
speed~\cite{Bertin,MarchettiPRE}. This ansatz reduces Eq.~\eqref{hydro1} to an
algebraic relation:
\begin{equation}
  \rho = \rho^\star + \frac{1}{c} W
\end{equation}
When a band moves in an isotropic gas, see e.g. Fig.~\ref{fig1}b, the
constant $\rho^\star$  corresponds to the gas density. 
Inserting the latter expression in Eq.~\eqref{hydro2} leads to a second order
ordinary differential equation:
\begin{equation}
\label{ODE}
D \ddot{W} + \dot W\frac{\d F}{\d W}  + \frac{\d H}{\d W} = 0,
\end{equation}
where $H(W)$ is defined via $dH/dW=a_2 W -a_4 W^3$,  $F(W) = \left(c-\frac{\lambda}{c} \right) W - \frac{1}{2} \xi W^2$, and the dot symbol denotes derivative with respect to $\tau \equiv x-ct$.
Therefore, the band-pattern problem is recast into a dynamical system framework: establishing the shape of the bands amounts to describing the motion of a  particle of mass $D$ and position $W$ in a potential 
$H(W)$, 
and experiencing a nonlinear friction $F(W)$, see Fig.~\ref{fig2}a. Note that the particle gains (resp. losses) energy when $F'(W)<0$ (resp. $F'(W)>0$).

\begin{figure*}
\begin{center}
\includegraphics[width=\textwidth]{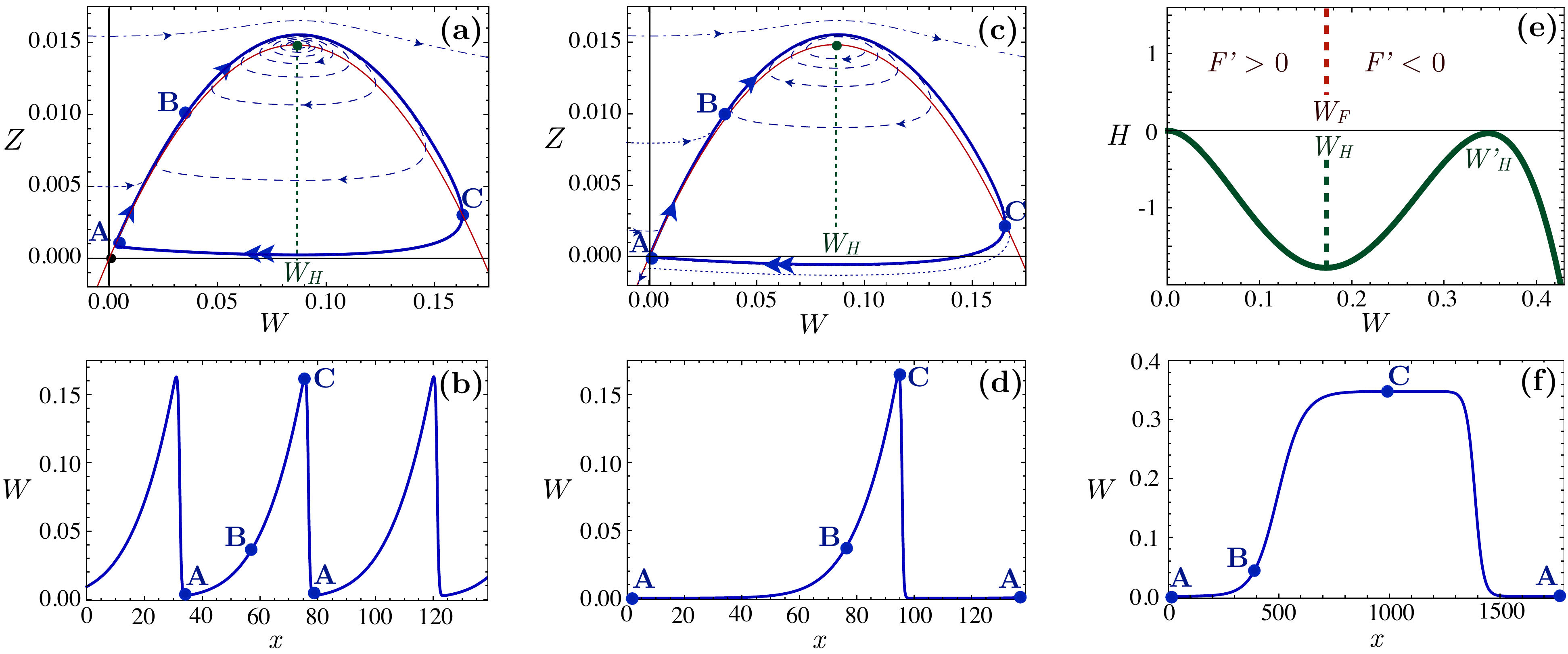}
\caption{(a) Dashed lines: dynamical system trajectories, for $P_0 =
  1$, $\rho_{\rm c} = 0.1$, $\lambda = 0.5$, $\xi = 4$, $D=0.1$,
  $c=0.9$. $\rho^\star$ was chosen such that $W_H \gtrsim W_F$. Thick
  line: stable limit cycle.  Thin line: $Z = F(W)$ curve. (b)~Polar smectic corresponding to the limit cycle shown
  in~(a). (c)~Homoclinic orbit; same parameters as in (a) but for a
  lower value of $\rho^\star$. (d) Solitonic band corresponding to the limit cycle shown
  in~(c). (e)~ $H(W)$, full line, plotted for $P_0 = 1$, $\rho_{\rm c} = 1$, $\lambda = 1$,
  $\xi = 10$ and $D=50$. The dashed lines show the positions of $W_F$ and $W_H$. The values of $c$ and $\rho^\star$ give rise
  to an heteroclinic cycle~\cite{supp}. (f)~Polar-liquid droplet for
  the same values of the parameters as in~(e), see also~\cite{supp}.}
\label{fig3}
\end{center}
\end{figure*}

Mass conservation in the original problem,
Eq.~\eqref{hydro1}, constrains the boundary conditions of
Eq.~\eqref{ODE} as $W(x\to -\infty) = W(x \to +\infty)$.  Given this
simple observation, without any further calculation, we can anticipate
all the possible band patterns: 
the solutions of Eq.~\eqref{ODE}
correspond to closed trajectories in the $(W,\dot W)$
plane. Therefore
they necessarily belong to one of the three following classes: (i)
periodic orbits, (ii) homoclinic cycles (the trajectory includes one
saddle point), or (iii) heteroclinic cycles (the trajectory includes
two saddle points). Back in real space, as exemplified in Fig.~\ref{fig3}, these trajectories respectively correspond to three possible propagating patterns $W(x-ct)$:
 (i) a smectic phase composed of ordered bands ($W$ varies
periodically with $x-ct$), (ii) a localized solitary wave, the length of
which being set by the "time'' taken to explore the homoclinic
cycle, (iii) a polar-liquid droplet separated by domain walls from an
isotropic gaseous phase, the fraction of polar liquid being given by
the ratio between the waiting times at the two saddle points. These
three patterns exactly correspond to those {\it hitherto} observed in
model experiments, and in numerical simulations at the onset of
collective motion.

Motivated by this pivotal observation, we now turn to the study of
equation~\eqref{ODE}. For sake of clarity we henceforth specify the
functional dependence of the phenomenological coefficients in
Eq.~\eqref{hydro2}. 
As the density is a control parameter of the
transition to collective motion for all models based on
short-range alignment
interactions, $a_2(\rho)$ has to
change sign at a finite density $\rho_{\rm c}$~\cite{Marchetti_review,Vicsek_review}. Different systems may result
in different functions $a_2(\rho)$. We choose a simple linear
dependence $a_2=\rho-\rho_{\rm c}$ which is consistent, close to $\rho_{\rm c}$,
with all existing
kinetic theories~\cite{Bertin,BaskaranPREmicro,TailleurXY,spins,rollers}.
In addition, having e.g. the original Vicsek model in mind, we want to capture the saturation of the average polarization of a homogeneous polar state of density $\rho=\rho_0$, when $\rho_0 \gg \rho_c$, at a non-zero value $P_0$. In agent-based models, $P_0$ is set by the noise amplitude. Here, since $P\sim \rho_0^{-1} [(\rho_0-\rho_c)/a_4]^{1/2}$, the simplest possible choice yielding the correction saturation is $a_4(\rho)=(\rho P_0^2)^{-1}$.
This choice simplifies the equations studied numerically in~\cite{MarchettiPRE}.  In all that follows, $\xi$, $\lambda$ and
$D$ are kept constant. 

Now that Eqs.~\eqref{hydro1}-\eqref{hydro2} have been fully defined, we can obtain their
propagative solutions explicitly by solving Eq.~\eqref{ODE}. 
We stress that the two functions $H(W)$ and $F(W)$ are parametrized by two independent parameters: $\rho^\star$ and $c$, which specify the shape of the bands. Their explicit form are provided in~\cite{supp} and $H$ is plotted in Fig.~\ref{fig2}a for a given set of parameters.
The existence of closed trajectories in the $(W,\dot W)$ 
plane requires that the system has at least
one fixed point~\cite{Arnoldbook}. Hence, keeping in mind that Eq.~\eqref{ODE} describes the motion of a massive particle in a potential, we look for trajectories that display at least one oscillation. This obviously requires: (i) that $H$ has a local minimum at a finite value $W_H > 0$ (and thus a local maximum at $W=0$), and (ii) that the friction $F'(W)$   changes sign at a finite $W_F>0$ so that  the particle do not fall to and remains stuck at $W_H$ where $H$ is minimum. It is straightforward to show that the former condition implies $\rho^\star<\rho_c$, and the latter  $c>\sqrt{\lambda}$. In order to establish the shape of the periodic trajectories of the dynamical system, and in turn the shape of the bands, we need to go beyond this simple picture.  We introduce the
auxiliary variable $Z \equiv D \dot W + F(W)$ and recast
Eq.~\eqref{ODE} into the 2-dimensional dynamical system:
\begin{align}
\label{DS1}	\dot W &= \frac{1}{D} \left[ Z - F(W) \right]\\
\label{DS2}	\dot Z &= - \frac{\d H}{\d W}
\end{align}
This change of variable greatly simplifies the investigation of the fixed points of the dynamical system now defined by Eqs~\eqref{DS1}-\eqref{DS2}~\cite{Strogatz}. 
It has at least two fixed points:
$(0,0)$ and $(W_H,F(W_H))$. A conventional linear stability analysis shows that $(0,0)$ is always a saddle point. Conversely
the second fixed point $(W_H,F(W_H))$ calls for a more careful
discussion. It undergoes a Hopf-bifurcation as $W_H - W_F$ changes
sign, as can be seen on the eigenvalues of the Jacobian
matrices~\cite{supp}.
This bifurcation, which we will thoroughly characterize
elsewhere~\cite{Solon2014}, is supercritical (resp. subcritical) if
$c<c^*$ (resp. $c>c^*$), where the critical velocity $c^*$ is defined
implicitly by $H'''(W_H)=0$. Both the bifurcation line and $c^*$ can
be computed analytically and are shown in Fig.~\ref{fig2}b. More
importantly, regardless of its sub- or super-critical nature the Hopf
bifurcation results in an unstable spiral trajectory which can lead
the system towards a cyclic attractor.  We now describe how these
limit cycles are explored in the $(W,Z)$ plane, and relate these
nonlinear trajectories to the morphologies of the band patterns.

{\em Polar smectic phase/periodic orbits.}  To gain more quantitative
insight, we consider large-amplitude cycles in the limit of small
$D$~\cite{Note1}. For small $W$, 
Eq.~\eqref{DS1} implies that the system quickly relaxes towards the curve $Z
= F(W)$ in a time $\sim D^{-1}$. Close to the origin, the dynamics is
controlled by the linear properties of the saddle point $(0,0)$, which
defines two well-separated scales. It can be easily shown that the
stable direction is nearly horizontal, it is associated with a fast
relaxation at the rate $\tau_-^ {-1}\sim D^{-1}(c - \lambda/c)
$. Conversely, the unstable direction is nearly tangent to the curve
$Z = F(W)$, and corresponds, again in the small-$D$ limit, to a much slower growth at the rate 
$\tau_+^ {-1}\sim  (\rho_{\rm c} - \rho^\star)/(c-\lambda/c)$.
The shape of large-amplitude cycles immediately follows from this
discussion and from the parabolic shape of $F(W)$.  Let us start from the left of the cycle, point~A in
Fig.~\ref{fig3}a, close to the origin. We call $W_{\rm min}$ the
abscissa of this point, which is the minimum value of $W$ in the
cycle. As noted above, the trajectory first remains near the parabola
$Z = F(W)$. If A is close enough to the origin, this part of the cycle
is explored slowly, in a time $\sim \tau_+$. Then the trajectory
approaches the unstable point $(W_H,F(W_H))$. It therefore leaves the parabola
 and starts spiraling, at a point labeled B in Fig.~\ref{fig3}a (B is here defined as the point where the trajectory deviates from the $Z=F(W)$ curve by 5\%).  It finally
crosses the parabola again, at point~C, and $\dot W$ changes sign, see Eq.~\eqref{DS1}. $W$ then decreases and the system quickly goes back to
point~A in a time typically set by $\sim\tau_-$. To further check this
picture we have numerically computed the phase portraits of Eqs.~\eqref{DS1} and~\eqref{DS2}, Fig.~\ref{fig3}a
(dotted lines). The typical periodic orbit shown in Fig.~\ref{fig3}a (full
line) is in excellent agreement with the scenario described above. From this analysis, 
we infer the shape of the steadily propagating band pattern $W(x-ct)$. As
anticipated, periodic orbits correspond to a polar smectic phase composed of equally-spaced  bands, in qualitative agreement with the experimental pictures reported in~\cite{Bausch}, and Fig.~\ref{fig1}a.
The numerical shape of  a smectic pattern is shown in
Fig.~\ref{fig3}b. It is composed of strongly asymmetric excitations,
which reflects the time-scale separation in the underlying dynamical
system close to the origin; the large-amplitude bands are composed of
a long exponential tail, and of a sharp front at the head. Note that
we describe here the propagation of  large-amplitude
excitations in a polarized environment. However the minimum
polarization in the regions separating the bands, $W_{\rm min}$, can
be vanishingly small. In this limit the period of the crystalline
structure would diverge logarithmically as $\tau_+\log \left(
W_F/W_{\rm min} \right)$.

{\em Solitary bands/Homoclinic cycles.}  
In the limiting case where the minimal value of $W$ goes to 0 ($W_{\rm min}=0$), point A then corresponds to the saddle located at the origin. Consequently the orbits followed by the dynamical system   become
homoclinic. As exemplified in Fig.~\ref{fig3}c, they are not periodic anymore, as the trajectory remains stuck at $(0,0)$.
In real space, the associated pattern is a solitary wave emerging out of a disordered gas, Fig.~\ref{fig3}d.
We stress that the existence of
solitonic structures at the onset of collective motion is one of the
more robust observations made in agent-based
simulations~\cite{ChatePRE,Bertin,Ihle2013}, see Fig.~\ref{fig1}b.

{\em Polar liquid droplets/Heteroclinic cycles.} Until now, we have
restricted our analysis to the case
where the dynamical system only probes the first two extrema of $H$.
However, looking for high-speed solutions ($c>P_0$), $H$ displays
an additional extremum at $W'_H>W_H$, see Fig.~\ref{fig3}e. $(W'_H,F(W'_H))$ is a second
saddle point. Therefore heteroclinic cycles are found when A is
located at the origin and C at the second saddle. The cycles are not periodic as the dynamics freezes both at $A$ and $C$.
In real space, the corresponding structure, $W(x-ct)$, is a localized domain, a polar-liquid droplet, traveling in a disordered gas, whose length is given by 
the residence at point C, Fig.~\ref{fig3}f.
This phase-separation pattern corresponds to the one numerically
observed in the active spin model,  Fig.~\ref{fig1}c, and in a generalization
of the 2D Toner and Tu model~\cite{spins,BaskaranPRE}. In the small $P_0$
limit, the shapes of the asymmetric domain walls bounding the
polar-liquid droplets can be computed exactly~\cite{supp}.

Several comments are in order. Firstly, we emphasize that the salient features of the swarming patterns do not
depend on the specific functional forms of the hydrodynamic  coefficients in Eq.~\eqref{hydro2}.  The 
limit-cycle solutions solely require the existence of a Hopf bifurcation, and
the dynamics along this cycle is chiefly controlled by the stability of the
 other fixed points. Therefore, at a
qualitative level, only the global shapes of the effective potential
$H(W)$ and the friction curve $F(W)$ matter. For instance, we shall stress that a hydrodynamic theory where $a_4=\rm cst$ in Eq.~\eqref{hydro2} would yield non-linear patterns qualitatively identical to those shown in Fig.~\ref{fig1} (not shown).  Only the sign reversal of $a_2(\rho)$ at $\rho_{\rm c}$ was necessary to observe band patterns, in agreement with~\cite{Peshkov}. 

Secondly, we  emphasize that travelling band  can exist when the average density $\rho_0$ is smaller  than $\rho_{\rm c}$, though the linear stability of Eqs. (1) and (2) predicts that no  small-amplitude wave can propagate~\cite{Marchetti_review}. The fundamental propagative excitations of polar active matter are intrinsically nonlinear below $\rho_c$.  

Thirdly, we come back to the status of the solutions described above. Until now we have
identified an infinite family of band-type solutions, 
located in the vicinity of the Hopf-bifurcation line in the $(c,\rho^\star)$ plane, grey region in
  Fig.~\ref{fig2}b. The boundaries of this region are found by looking for non-degenerate solutions satisfying $W>0$~\cite{Solon2014}, and its  extent 
 is an increasing function of the diffusivity $D$. The domain of existence of the bands
collapses on the Hopf-bifurcation line in the limit $D\to0$. The homoclinic cycles, corresponding to solitary waves, are constraint to include one saddle point. Therefore they define a one-parameter
ensemble of band-type solutions. This ensemble corresponds to the lower boundary of the phase diagram,  Fig.~\ref{fig2}b (dashed-dotted line), and is  established by taking the infinite-period limit. The heteroclinic solution, polar-liquid droplet, is constrained by the existence of two saddles along a cycle. Therefore, if any, the heteroclinic cycle is unique. It is  a limiting case of the one-parameter homoclinic family, point $c_{\rm h}$ in Fig.~\ref{fig2}b.  

Finally, we discuss the pattern-selection problem. The ensemble of band-type solutions described above is actually restrained by the mass-conservation law.
The mean density $\rho_0$ in the system is fixed,
therefore only band shapes compatible with this value exist.   However, we are
a priori left with an infinite family of solutions, which is  parametrized by one free parameter.
Hence, we predict that, for a given value of $\rho_0$, several
solutions propagating at different speeds may coexist. This conjecture
is again supported by numerical evidences. In Fig.~\ref{fig1} the
three-band and single-band patterns correspond to identical values of
all the simulation parameters. The  full resolution of the challenging pattern-selection problem obviously goes beyond the scope of this letter. However, a tentative picture
for the nucleation of stationary swarms from a disordered state can be attempted 
from Eq.~\eqref{hydro2}. 
The emergence of sharp fronts is natural since the  l.h.s. of~\eqref{hydro2} has the form of Burgers equation, 
which supports rarefaction shocks~\cite{Whitham}. A density fluctuation above $\rho_{\rm c}$ grows and polarizes
via the generic coupling between density and order embodied in the $\rho$-dependence of $a_2$ in Eq.~\eqref{hydro2}. 
When these two competing effects balance each other, the
density at the top of the shock is pinned, and a constant-shape asymmetric band steadily propagates.
In the transient regime, we therefore expect several bands to form and collide, until the system
reaches one of the possible steady states. This mechanism might favor
large-amplitude/fast bands via coalescence events, in agreement with
the experiments reported in~\cite{rollers}.

To close this letter we comment on  the role of fluctuations on the transition towards collective
motion~\cite{Vicsek_review}. Eq.~\eqref{hydro2} predicts a second-order transition  for  homogeneous systems. Here, we have evidenced that stationary
polarized excitations (solitary bands, and polar-liquid droplets)
can coexist with a homogeneous isotropic phase, which in turn confirms the first order
scenario evidenced in numerical simulations~\cite{ChatePRL,ChatePRE}. 
This coexistence does not rely on any fluctuation-induced mechanism, unlike all the  conventional equilibrium scenarios making first order a mean-field second-order  transition (e.g. Brazowskii~\cite{Brazowskii} and Halperin-Lubensky-Ma~\cite{HLM}). However, beyond the mean-field deterministic picture, fluctuations are very likely to play a major role in the stability, the selection, and the ordering of the band patterns. These difficult but crucial problems are the topic for future work.

\begin{acknowledgments}
We thank J. Toner for valuable comments and suggestions.
DB acknowledges support from Institut Universitaire de France, and ANR project MiTra. HC, AP, AS, and JT thank
the Max Planck Institute for the Physics of Complex Systems, Dresden,
for providing the framework of the Advanced Study Group ``Statistical Physics of Collective Motion'' within which part of this work was performed.\end{acknowledgments}

\balancecolsandclearpage

\onecolumngrid
\appendix

\titleformat{\section}{\center \small \bfseries}{\thesection. }{0pt}{\MakeTextUppercase}   \titlespacing*{\section}{0pt}{6ex plus 1ex minus .2ex}{4ex plus .2ex}
\titleformat{\subsection}{\center \small \bfseries}{\thesubsection. }{0pt}{}    \titlespacing*{\subsection} {0pt}{3.25ex plus 1ex minus .2ex}{3ex plus .2ex}

\begin{center}
	\large{\bf Supplementary Information}
\end{center}

\section{I. Microscopic models}
Figure 1 of the main text shows typical examples of inhomogeneous
structures found in polar flocking models. Fig.~1a and 1b stem from a
Vicsek model with vectorial noise~\cite{ChatePRE} while Fig. 1c comes
from an active Ising model \cite{Solon}.  We recall below the definition of these models, while more details can be found
in the original publications~\cite{ChatePRE,Solon}.

\subsection{A. Vicsek model}
$N$ point-like particles are moving off-lattice in 2d at constant speed
$v_0$. At each time step, the direction along which each particle moves
is updated in parallel, according to the rule
\begin{equation}
  \label{eq:vicsek_angular}
  \theta_i(t+1)=\arg \left(\sum_{j\in \mathcal{S}_i} \vec{v_j}(t)+\eta \mathcal{N}_i\vec{\xi}\right)
\end{equation}
where $\theta_i$ is the angle made by the velocity $\bf v_i$ of the $i^{\rm th}$ particle and, say the $x$-axis. The sum runs on particles in
$\mathcal{S}_i$, the neighborhood of $i$ of radius $r_0$. $\mathcal{N}_i$ is the number of particles in $\mathcal S_i$ and
$\vec{\xi}$ is a random unit vector with no angular correlation.   This equation defines 
so-called vectorial noise version of the Vicsek model~\cite{Vicsek,ChatePRE}.  
\subsection{B. Active Ising model}
$N$ particles carrying a spin $\pm 1$ move and interact on a 2d
lattice, an arbitrary number of particles being allowed on each
site. The particles can hop to any adjacent sites or flip theirs spins
according to Monte-Carlo rates. The spin-flipping results from an
on-site ferromagnetic interaction, a particle reverses its spin $S$ on site $i$ at rate
\begin{equation}
  \label{eq:ising_interaction}
  W(S\to -S)=\exp\left(-\beta S \frac{m_i}{\rho_i}\right)
\end{equation}
where $m_i$ and $\rho_i$ are the magnetization and the number of
particles on site $i$ and $\beta$ plays the role of an inverse
temperature.
The ``self-propulsion'' of the particles stems from a bias on the
hopping direction: particle $k$ hop to the left (resp. right) with
rates $D(1+\varepsilon S_k)$ (resp. $D(1-\varepsilon S_k)$) and to the
top and bottom with rate $D$. This sets an effective self-propulsion
at speed $2D\varepsilon S_k$ in the horizontal direction.

\section{II. Expressions of $H$ and $F$}

Using the functional dependences of $a_2(\rho)$ and $a_4(\rho)$ that
we introduced in the main text, we find the following expressions for
the friction $F$ and the potential $H$:
\begin{align}
	&F(W) = \left(c-\lambda c^{-1}  \right) W - \frac{1}{2} \xi W^2,\\
	&\frac{\d H}{\d W} = \left[ -(\rho_c - \rho^\star) + \frac{W}{c} - \frac{c W^2}{P_0^2 (c \rho^\star + W)} \right] W.
\end{align}
The latter equation can be integrated over $W$ and yields
\begin{align}
	H(W) =& -\frac{1}{2} (\rho_c - \rho^\star) W^2 + \frac{1}{3c} W^3 - \frac{c^3 {\rho^*}^2}{P_0^2} W + \frac{c^2 \rho^*}{2P_0^2} W^2 \nonumber \\& \quad- \frac{c}{3P_0^2} W^3 + \frac{c^4 {\rho^*}^3} {P_0^2} \log \left( 1 + \frac{W}{c \rho^\star} \right).
\end{align}
\section{III. Eigenvalues of the Jacobian matrix}
\begin{figure}[h]
\begin{center}
\includegraphics[scale=0.45]{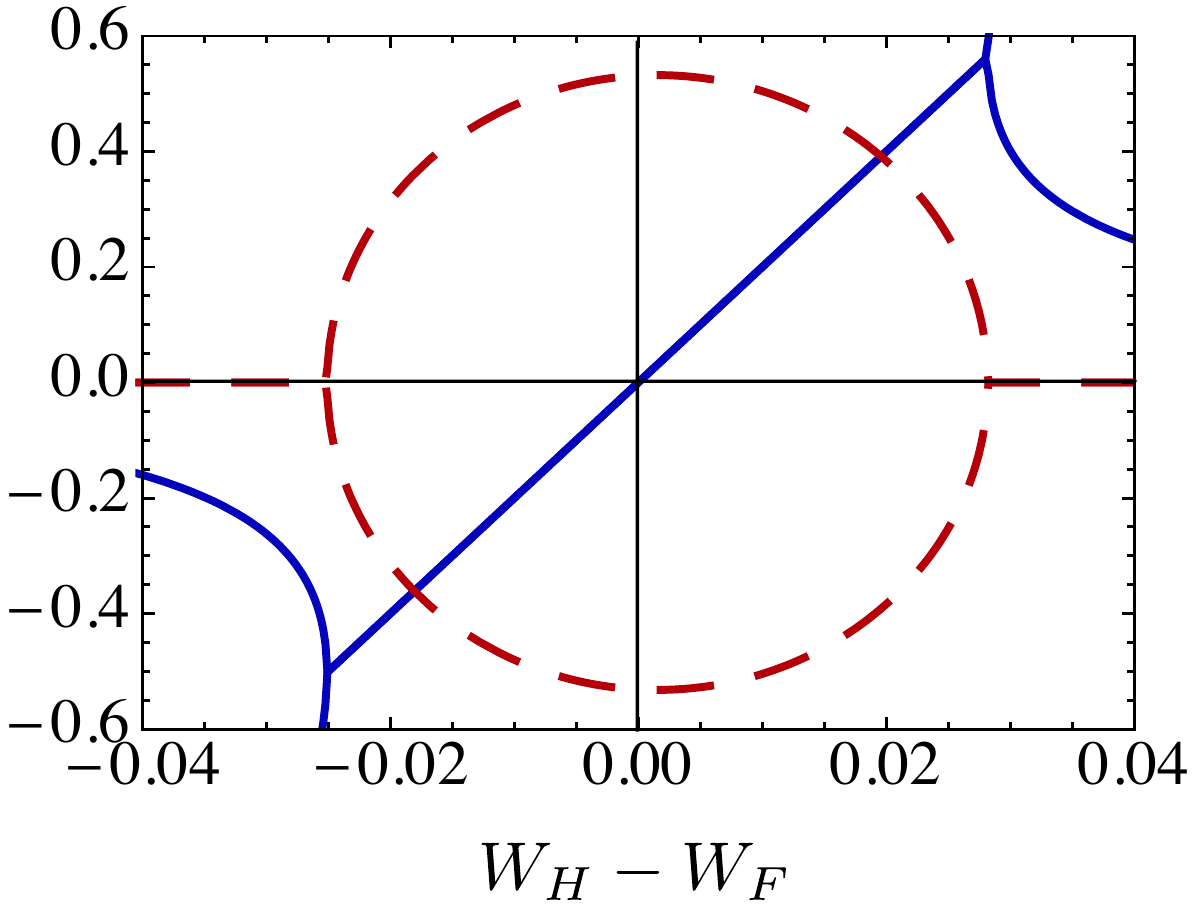}
\caption{Eigenvalues of the Jacobian matrix at the fixed point $(W_H,F(W_H))$, as a function of $W_H-W_F$. The curves were plotted for fixed $c = 0.9$ by varying $\rho^\star$. Blue solid line: real part. Red dashed line: imaginary part.}
\label{eigenvalues}
\end{center}
\end{figure}

The eigenvalues of the Jacobian matrix evaluated at the fixed point $(W_H,F(W_H))$ are plotted in Fig.~\ref{eigenvalues}. The signs of their real and imaginary parts are only set by the difference $W_H - W_F$, hence a Hopf bifurcation occurs at $W_H = W_F$.

\section{IV. Polar liquid domains/Heteroclinic cycles}
\begin{figure}[h]
\begin{center}
\includegraphics[scale=0.8]{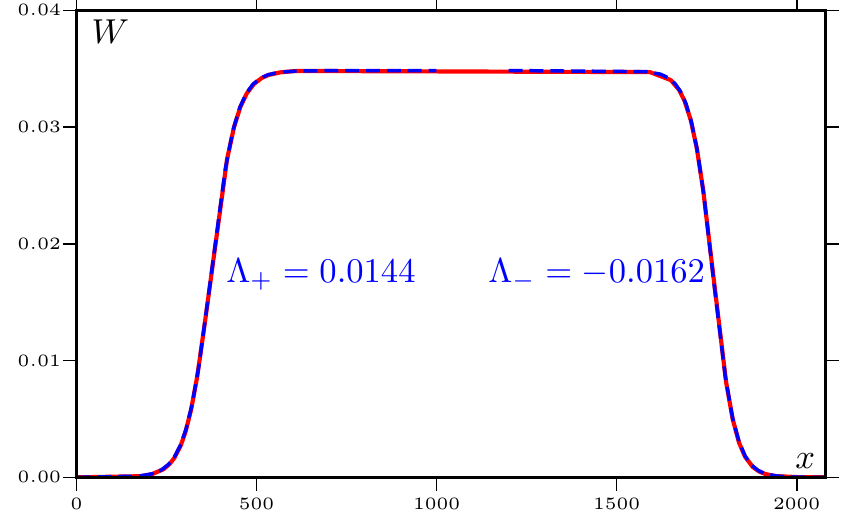}
\caption{Heteroclinic orbit obtained by numerical integration (plain
  red line) and fit with Ansatz (\ref{profile_wall}) (dashed blue
  lines). Fitting parameters are the width of the front $\Lambda_\pm$
  and the position of the front.  Parameters: $P_0=0.1$, $\rho_c=1$,
  $\xi=10$, $\lambda=0.01$, $D=50$, $c=0.218$, $\rho^*=0.9502$.}
\label{hetero}
\end{center}
\end{figure}

We compute analytically the shape of the heteroclinic orbits, {\it i.e.}
the boundary between a  polar-liquid droplet and the surrounding  isotropic gas, in the limit of small $P_0$.
Since the magnetization scale is set by $P_0$, when $P_0\to 0$, $W\ll
\rho^*$ so that the Hamiltonian reduces to
\begin{equation}
  \label{eq:dH_cubic}
  \frac{dH}{dW}=\left[-(\rho_c-\rho^*)+\frac{W}{c}-\frac{W^2}{P_0^2\rho^*}\right]W
\end{equation}
Note that this cubic expression becomes exact if we assume the coefficient $a_4$ to be independent of $\rho$ and $W^2$, in the hydrodynamic equations (Eq.~(2) main text).
With this potential, solutions can be found using the Ansatz
\begin{equation}
\label{profile_wall}
	W_\pm(x - ct) = \frac{W_H'}{2} \left[ 1 + \tanh \left( \Lambda_\pm (x-ct) \right) \right] ,
\end{equation}
which yields excellent agreement with the numerical solution of the ODE, see Fig.~\ref{hetero}.
For a given set of parameters, only one heteroclinic cycle is found in
the ($\rho^*$, $c$) plane and indeed, inserting
Eq.~(\ref{profile_wall}) in the ODE with the potential given by
Eq.~(\ref{eq:dH_cubic}) we obtain a unique solution ($\rho^*$,
$c$) given by
\begin{align}
  \label{eq:ch}
  c&=\frac{1}{3\sqrt{2}} \sqrt{\tilde P_0+9 \lambda +\sqrt{72 P_0^2 \lambda + (9 \lambda + \tilde P_0^2)^2}} \\
  \rho^*&=\frac{9c^2\rho_c}{9c^2+2 P_0^2} \\
\label{WH'}	W_H' &= \frac{2 P_0^2 \rho^\star}{3c} \\
  \Lambda_\pm &=\frac{(c-\frac{\lambda }{c})}{4D} \left (-1 \pm \sqrt{1+\frac{4 D (\rho_c-\rho^*)}{(c-\frac{\lambda }{c})^2}}\right)
\end{align}
where $\tilde P_0=P_0^2(3\xi\rho_c-2)$.
Table~\ref{tab:comparison} compares the five parameters $c$, $\rho^\star$, $W_H'$, $\Lambda_+$ and  $\Lambda_-$ found analytically to the best fit. The agreement is excellent.
\begin{table}[h]
  \centering
  \begin{tabular}{|c|c|c|}
    \hline
    & Numerical & Analytical \\\hline\hline
   c & 0.218 & 0.2041 \\\hline
   $\rho^*$ & 0.9502 & 0.949 \\\hline
   $W_H'$ & 0.0348 & 0.0310 \\\hline
   $\Lambda_+$ & 0.0144  & 0.0152 \\\hline
   $\Lambda_-$ &  -0.0162 & -0.0167 \\\hline
 \end{tabular}
 \caption{Comparison between numerical and analytical values obtained in the limit $P_0\to 0$. Parameters: $P_0=0.1$, $\rho_c=1$, $\xi=10$, $\lambda=0.01$, $D=50$. Numerically, $\Lambda_\pm$ are obtained
   by fitting the profile with Ansatz (\ref{profile_wall}).}
  \label{tab:comparison}
\end{table}\\

\begin{acknowledgments}
We thank J. Toner for a careful reading of the supplementary document and for suggesting the compact formula shown in Eq.~\eqref{WH'}.
\end{acknowledgments}

\end{document}